# CONCEPTUAL DESIGN OF A 20 T DIPOLE BASED ON HYBRID REBCO/Nb$_3$Sn COS-THETA COIL*

A. V. Zlobin†, Fermilab, Batavia, IL 60510, USA

*Abstract*

This paper presents a design concept of an HTS/LTS hybrid dipole with 50 mm aperture and 20 T nominal field based on a cos-theta coil and a cold iron yoke. The HTS part of magnet coil uses REBCO Twisted Stacked Tape cable. The LTS part is graded and made of two Nb$_3$Sn Rutherford cables. Due to high sensitivity of HTS/LTS coils to large stresses and strains at high fields, a Stress Management (SM) concept combined with the cos-theta coil geometry is used. The results of magnet magnetic analysis are presented and discussed.

## INTRODUCTION

Dipole magnets with a nominal operation field up to 20 T are being considered for the next generation of particle accelerators. Since the field level of 20 T is above the practical limit of Nb$_3$Sn accelerator magnets, it requires using High Temperature Superconductors (HTS) with higher upper critical magnetic field $B_{c2}$ and critical temperature $T_c$. Due to the high cost of HTS materials and the more complex technology of HTS magnets, a hybrid design, based on both HTS and Nb$_3$Sn cables and technologies, is the most optimal to minimize the volume and the cost of HTS coil and the magnet in general. In a hybrid design, HTS cables are used in the high-field coil area and Nb$_3$Sn cables are used in the outer, lower-field coil areas.

Various design options of a 50-mm aperture 20 T dipole are being studied in the framework of US-MDP [1]. The studies include Cos-Theta (CT), Block (BL) and Common-Coil (CC) coil configurations [2]. Complementary studies of HTS and hybrid HTS/LTS dipole magnets are also being performed at Fermilab. A 20 T superconducting (SC) dipole based on hybrid Bi2212 and Nb$_3$Sn coils with stress management has been reported in [3].

This paper describes a design concept of a hybrid dipole magnet with 50 mm aperture and 20 T nominal field based on a REBCO/Nb$_3$Sn CT coil and a cold iron yoke. The results of magnet magnetic design optimization and analysis are presented and discussed. The key parameters of this design are compared with the parameters of magnet design based on hybrid Bi2212/Nb$_3$Sn dipole coil [3].

## MAGNETIC DESIGN AND PARAMETERS

The magnetic design and analysis has been performed using ROXIE code [4] with real iron $B(H)$ curve. The hybrid coil design is based on Twisted Stacked Tape cable made of composite REBCO (HTS) tapes and two Rutherford cables made of Nb$_3$Sn (LTS) composite SC wires.

* Work supported by Fermi Research Alliance, LLC, under contract No. DE-AC02-07CH11359 with the U.S. DOE and by the U.S. Magnet Development Program (US-MDP).
† zlobin@fnal.gov

Main parameters of the cables and the cable insulation thickness are shown in Table 1.

The Nb$_3$Sn Rutherford cables use round strands and have keystoned cross-sections represented by the width and the thickness of small and large edges.

A picture of Twisted Stacked Tape is shown in Fig. 1 [5]. In ROXIE this cable has been represented by the rectangular single-strand cable with parameters shown in Fig. 2. The REBCO cable is a square stack of 50 parallel 5-mm wide 0.1-mm thick tapes twisted in the coil straight section along the stack axis during coil winding. The orientation of tapes in the stack at both ends of the coil straight section is radial to ensure cable "easy" bend around narrow poles. The possibilities and limitations of REBCO tape twisting and bending are reported in [5].

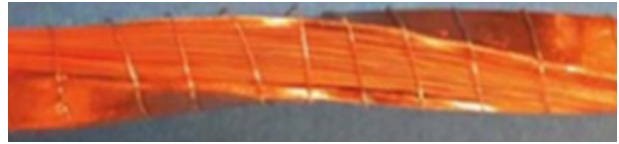

Figure 1: A view of a Twisted Stacked Tape cable [5].

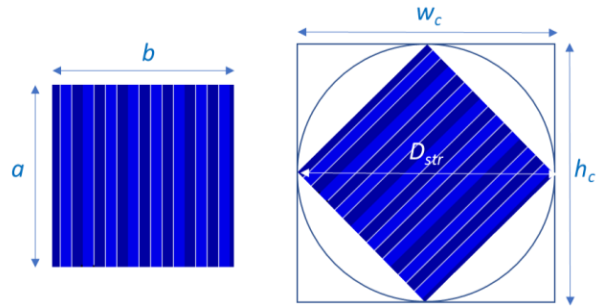

Figure 2: Cross-section parameters of the REBCO TST cable (left) and its ROXIE representation (rigth): $a$ – tape stack width; $b$ – tape stack thickness = tape thickness × number of tapes; $D_{str}$ – equivalent diameter of the REBCO stacked tape strand; $w_c$, $h_c$ – rectangular cable width and thickness respectively.

Table 1: Strand and Cable Geometrical Parameters

| Parameter | Cable 1 | Cable 2 | Cable 3 |
|---|---|---|---|
| Superconductor | REBCO | Nb$_3$Sn | Nb$_3$Sn |
| Strand size $D_{str}$, mm | 7.1 | 1.1 | 0.8 |
| Cu/nonCu ratio | ~0.25* | 1.0 | 1.1 |
| Number of strands | 1 | 38 | 38 |
| Cable width $w_c$, mm | 7.1 | 21.5 | 15.7 |
| Cable small edge $h_c$, mm | 7.1 | 1.85 | 1.39 |
| Cable large edge $h_c$, mm | 7.1 | 2.05 | 1.58 |
| Cable packing factor** | ~0.5 | 0.90 | 0.85 |
| Insulation thickness, mm | 0.45 | 0.15 | 0.15 |

* Cu/nonCu ratio for REBCO cable is a ratio of Cu cross-section in tapes to the strand cross-section
** cable packing factor is defined as a ratio of the total strand/tape cross-section to the cable cross-section.

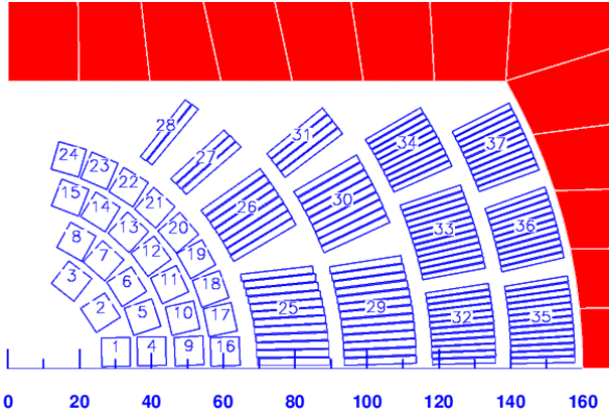

Figure 3: Cross-section of one quadrant of the 20 T hybrid REBCO/Nb$_3$Sn dipole coil inside the iron yoke.

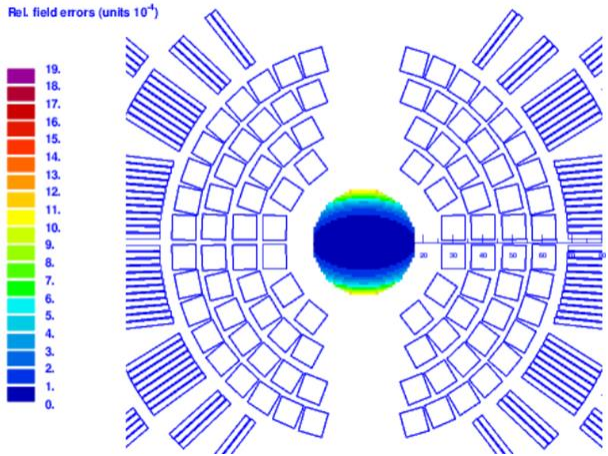

Figure 4: Field quality diagram in the aperture within the 17-mm radius circle at a bore field of 20 T.

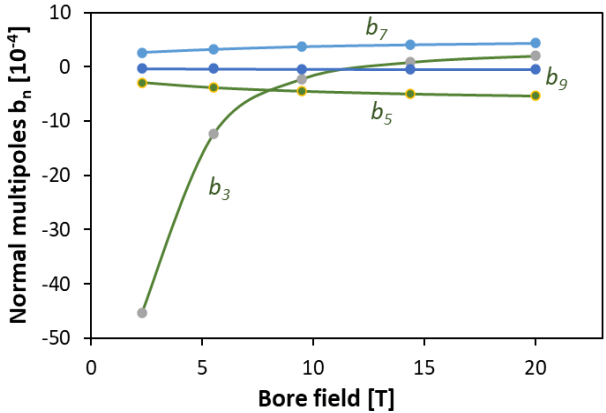

Figure 5: Low-order field harmonics $b_3$, $b_5$, $b_7$ and $b_9$ vs. the bore field $B_o$.

Table 2: Geometrical Field Harmonics at $R_{ref}$=17 mm.

| $n$ | 3 | 5 | 7 | 9 |
|---|---|---|---|---|
| $b_n$, $10^{-4}$ | 1.97 | -5.37 | 4.40 | -0.49 |

The optimized cross-section of the 20 T hybrid REBCO/Nb$_3$Sn dipole coil is shown in Fig. 3. The coil consists of 4 double-layer coils with 24 HTS and 13 Nb$_3$Sn blocks. The number of turns in L1-2/L3-4/L5-6/L7-8 is 8/16/51/70 respectively. Coil blocks are numbered starting from the midplane towards the poles as shown in Fig. 3. The coil inner and outer diameters are 50 mm and 315 mm respectively.

Individual turns of the REBCO coil are separated in azimuthal and radial directions. Turns in the Nb$_3$Sn coils are combined into blocks divided by radial (interlayer) and azimuthal spacers. The radial and azimuthal space between turns and blocks is used to optimize the field quality in the aperture. It will be used also to control the coil geometry and the mechanical stress in the coil by placing coil turns or blocks in grooves in special coil mandrel [6]-[8].

The iron cross-section in this design is the same as in the 20 T Bi2212/Nb$_3$Sn hybrid dipole [3]. The outer radius of the iron yoke is 75 cm, the inner yoke shape is shown in Fig. 3. The yoke horizontal surface is separated from the midplane by 80 mm and the vertical surface has a radius of 160 mm. The effect of the yoke inner shape and outer diameter (OD) on the value of magnetic field in the coil and in the aperture, as well as on the geometrical low-order harmonics is discussed in [3].

Geometrical field harmonics were optimized in the first approximation for the nominal field of 20 T. Table 2 summarizes the low-order geometrical harmonics in the magnet aperture at the reference radius $R_{ref}$ = 17 mm. The field quality diagram (dark-blue area) inside a circle with a radius of 17 mm in the magnet aperture at $B_o$=20 T is plotted in Fig. 4. The good field area where the relative field error is less than $3 \cdot 10^{-4}$ has elliptical shape with the vertical and horizontal axis of approximately 20 mm and 40 mm.

Figure 5 shows the effect of the bore field $B_o$ on the low-order field harmonics $b_3$, $b_5$, $b_7$ and $b_9$. The iron saturation leads to large variations of $b_3$ vs. $B_o$ whereas its effect on the higher-order harmonics $b_5$, $b_7$ and $b_9$ is practically negligible. The final optimization of field quality in magnet aperture in the operation field range will be done by optimizing cross-sections of the coil and the iron and will include the superconductor magnetization.

Figure 6 shows the calculated distribution of magnetic field in the coil at the nominal bore field of 20 T. The maximum field level and, thus, the minimal margins of two REBCO (L1-2 and L3-4) and two Nb$_3$Sn (L5-6 and L7-8) coils are determined by the pole blocks 3 (L1), 15 (L3), 28 (L5) and 34 (L7).

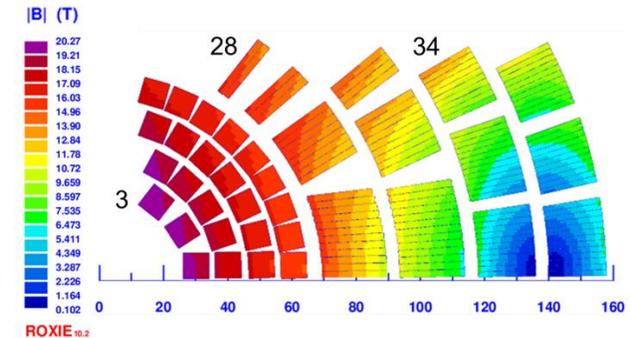

Figure 6: Calculated distribution of the magnetic field in the coil at $B_0$=20 T. The numbers show the critical coil blocks in two REBCO and two Nb$_3$Sn double-layer coils.

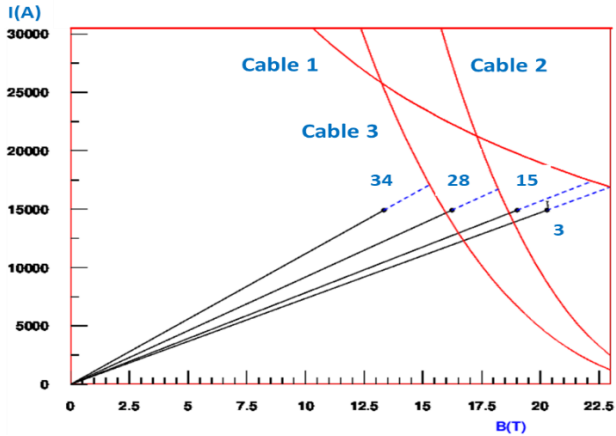

Figure 7: $I_c(B)$ curves at 1.9 K of the three cables used in the dipole, and the load lines of the key coil blocks.

Figure 7 shows the $I_c(B)$ curves of REBCO TST Cable 1 and Nb$_3$Sn Cables 2 and 3 at 1.9 K, and the maximum load lines for pole block 3 of the REBCO coils, and pole blocks 28 and 34 of the Nb$_3$Sn coils. REBCO cable parameters correspond to the tape reference $I_c$ at 15 T and 4.2 K of 95 A per mm of the tape width in magnetic field perpendicular to the tape surface. Parameters of both Nb$_3$Sn cables corresponds to Nb$_3$Sn strand $J_c$ at 15 T and 4.2 K of 1.2 kA/mm$^2$. The parameters of all the cables used in the described design are presently achieved in the commercial REBCO tapes [9, 10] and composite Nb$_3$Sn strands [11].

The minimal margins to quench along the load line for the three coil grades are 11.8% for REBCO Coil 1 (L1-2), 10.9% for Nb$_3$Sn Coil 3 (L5-6), and 12.8% for Nb$_3$Sn Coil 4 (L7-8). The magnet quench margin is determined by the Nb$_3$Sn coil 3 (L5) and is 10.9%. Various possibilities of increasing the quench margin of REBCO and Nb$_3$Sn coils are being studied and will be reported separately. They include optimization of superconductor fraction in cables, superconductor $J_c$, cable grading, etc.

## COMPARISON OF HYBRID DIPOLES WITH REBCO AND BI2212 HTS COILS

Cross-sections of REBCO/Nb$_3$Sn and Bi2212/Nb$_3$Sn hybrid dipole coils developed at Fermilab inside the iron yoke in the same scale are shown in Fig. 8. To fit the REBCO/Nb$_3$Sn coil into the same space in the yoke, the thickness of the interlayer spacer in Nb$_3$Sn coils was reduced from 5 mm to 2 mm due to wider Nb$_3$Sn cables. It reduces the effectiveness of stress management for Nb$_3$Sn coils. The mechanical stress level and distribution in both magnet designs will be further studied and optimized.

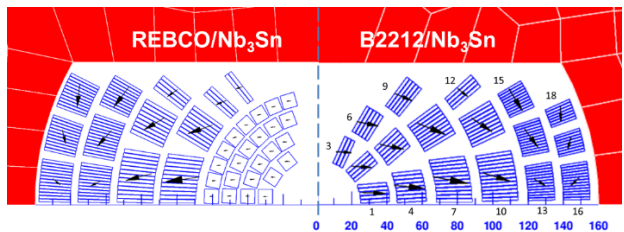

Figure 8: Cross-sections of HTS/Nb$_3$Sn hybrid dipole coils with Lorentz force vectors in blocks inside the iron yoke.

Table 3: Magnet Parameters

| Parameter | REBCO/ Nb$_3$Sn | Bi2212/ Nb$_3$Sn |
|---|---|---|
| Nominal current $I_{nom}$, kA | 14.90 | 13.45 |
| Nominal bore field $B_o$, T | 20.0 | 20.0 |
| Coil nominal field $B_{nom}$, T | 20.34 | 20.04 |
| Magnet $TF=B_o/I_{nom}$ at $I_{nom}$, T/kA | 1.342 | 1.487 |
| Magnet margin at 1.9 K, % | 10.9 | 13.2 |
| Total HTS coil area, mm$^2$ | 601 | 972 |
| Total Nb$_3$Sn coil area, mm$^2$ | 3854 | 3110 |
| Total coil area, mm$^2$ | 4455 | 4082 |

The main parameters of both magnets are summarized in Table 3. The REBCO/Nb$_3$Sn dipole design provides the target nominal field of 20 T at higher current with 21% lower magnet quench margin. However, the HTS coil cross-section in this design is ~30% smaller whereas the total coil cross-section is ~10% larger than in Bi2212/Nb$_3$Sn dipole design.

## CONCLUSION

Conceptual design of a 20 T hybrid dipole demonstrator based on REBCO and Nb$_3$Sn coils has been developed and analyzed. The HTS part of magnet coil in this design uses a twisted rectangular stack of REBCO tape or so called Twisted Stacked Tape cable. The LTS part is graded and made of two Nb$_3$Sn Rutherford cables. The twist of stacked tape cable along the stack axis is provided in the coil straight section during coil winding. This technology is being developed at Fermilab using sub-scale dipole models.

The magnet provides a nominal target field of 20 T with ~11% quench margins at 1.9 K with the state-of-the-art superconductor parameters, four double-layer hybrid shell-type graded coils and cold iron yoke. The magnet quench margin is determined by the Nb$_3$Sn coils. Possibilities of increasing the Nb$_3$Sn coil and total magnet margin are being studied.

The cross-section of the HTS coil in the presented REBCO/Nb$_3$Sn hybrid dipole is ~30% smaller with respect to the Bi2212/Nb$_3$Sn hybrid design described in [3] whereas the total coil cross-section is ~10% larger.

As in the Bi2212/Nb$_3$Sn hybrid dipole [3], the stress management elements in the REBCO/Nb$_3$Sn hybrid dipole are integrated into the coil cross-section to keep the mechanical stresses in brittle REBCO and Nb$_3$Sn superconductors below their limits. The analysis and optimization of the coil stress management structure as well as the magnet quench protection studies will follow.